\documentstyle[aps,prb,epsfig,eqsecnum,%
multicol,amsfonts,amsbsy,texdraw,times]{revtex}
\tighten
\draft
\begin{document}
\title{Critical behavior at $\boldsymbol{m}$-axial Lifshitz points:
field-theory analysis and $\boldsymbol{\epsilon}$-expansion results }
\author{H.\ W.\ Diehl}
\address{Fachbereich Physik, Universit\"at - Gesamthochschule  Essen,
D-45117 Essen, Federal Republic of Germany$^\ddag$ and}
\address{Physics Department, Virginia Polytechnic Institute
and State University, Blacksburg, VA 24601, USA}
\author{M.\ Shpot$^\dag$\thanks{email: shpot@theo-phys.uni-essen.de}}
\address{Fachbereich Physik, Universit\"at - Gesamthochschule  Essen,
D-45117 Essen, Federal Republic of Germany}
\date{\today}
\maketitle
\begin{abstract}
The critical behavior of $d$-dimensional systems
with an $n$-component
order parameter is reconsidered
at $(m,d,n)$-Lifshitz points, where a wave-vector instability occurs
in an $m$-dimensional subspace of ${\mathbb R}^d$.
Our aim is to sort out which ones of the previously published partly contradictory
$\epsilon$-expansion results to second order in
$\epsilon=4+\frac{m}{2}-d$ are correct.
To this end, a field-theory calculation is performed directly
in the position space of $d=4+\frac{m}{2}-\epsilon$ dimensions,
using dimensional regularization and minimal subtraction of
ultraviolet poles.
The residua of the dimensionally
regularized integrals that are required to
determine the series expansions of the correlation exponents
$\eta_{l2}$ and $\eta_{l4}$
and of the wave-vector exponent $\beta_q$
to order $\epsilon^2$ are reduced to
single integrals, which for general $m=1,\ldots,d-1$
can be computed numerically,
and for special values of $m$,  analytically.
Our results are at variance with the original
predictions for general $m$.
For  $m=2$ and $m=6$, we confirm the results of
Sak and Grest [Phys.\ Rev.\ B {\bf 17}, 3602 (1978)] and
Mergulh{\~a}o and Carneiro's recent field-theory analysis
[Phys.\ Rev.\ B {\bf 59},13954 (1999)].
\end{abstract}
\begin{multicols}{2}
\section{Introduction}
A Lifshitz point \cite{HLS75b,HLS75,Hor80,Sel92}
is a critical point at which
a disordered phase, a spatially homogeneous ordered phase,
and a spatially modulated phase meet.
In the case of  a $d$-dimensional system
with an $n$-component order parameter,
it is called an $(m,d,n)$-Lifshitz point
(or $m$-axial Lifshitz point)
if   a wave-vector instability occurs in an $m$-dimensional subspace.
Such multi-phase points are known to occur in a variety
of distinct physical systems,
including magnetic ones \cite{SBOC81,FurRef}, ferroelectric crystals \cite{VS92},
charge-transfer salts,\cite{AD77,HZW80} liquid crystals,\cite{CL76}
systems undergoing structural phase transitions\cite{AM80} or
having domain-wall instabilities,\cite{LN79} and the
ANNNI model.\cite{FS80,FS81} A survey covering the work related to them
till 1992 has been given by Selke,\cite{Sel92} which complements and updates
an earlier review by Hornreich.\cite{Hor80}
Recently there has also been renewed interest in the effects
of surfaces on the critical behavior at Lifshitz points.\cite{Gum86,BF99,FKB00}

From a general vantage point, critical behavior at Lifshitz points
is an interesting subject in that it presents clear and simple examples of
\emph{anisotropic scale invariance}. Epitomized also
by dynamic critical phenomena near thermal equilibrium,\cite{HH77}
and known to occur as well in other static equilibrium systems (e.g.,
uniaxial dipolar ferromagnets),
this kind of invariance has gained increasing attention
in recent years since it was found to be realized in many
non-equilibrium systems such as driven diffusive systems \cite{SZ95}
and in growth processes.\cite{Kru97}

Systems at Lifshitz points are good candidates for studying
general aspects of anisotropic scale
invariance.\cite{ASIERW,Hen97}
For one thing, the continuum theories representing
the universality classes of systems
with short-range interactions at $(m,d,n)$-Lifshitz points
are conceptually simple;
second, they involve the degeneracy $m$
as a parameter, which can easily be varied between $1$ and $d$.
A thorough understanding of critical behavior at such Lifshitz
points is clearly very desirable.

The problem has been studied decades ago by means of
an $\epsilon$ expansion about the upper
critical dimension\cite{HLS75b,Muk77,HB78,SG78}
\begin{equation}
d^*(m)=4+\frac{m}{2}\;,\quad m\le 8\,.
\end{equation}
Other investigations employed
the dimensionality expansion about
the lower critical dimension\cite{GS78}
 $d_*(m)=2+\frac{m}{2}$ for  $n\ge 3$,
or the $1/n$ expansion. \cite{HLS75,NTCS76,IHB90}
Unfortunately, the $\epsilon$-expansion results
to order $\epsilon^2$
one group of authors\cite{HLS75b,Muk77,HB78}
obtained for the correlation exponents
$\eta_{l2}$ and $\eta_{l4}$ and
the wave-vector exponent $\beta_q$ are in conflict
with those of Sak and Grest\cite{SG78} for
the cases $m=2$ and $m=6$.

In order to resolve this
long-standing controversy,
Mergulh{\~a}o and Carneiro\cite{MC98,MC99}
recently presented a reanalysis of the problem based
on renormalized field theory and dimensional regularization.
Exploiting the form of the resulting renormalization-group equations,
they were able to derive various
(previously given) general scaling laws
one expects to hold according to the phenomenological theory of
scaling. However, their calculation
of critical exponents
was limited in a twofold fashion:
They treated merely the special cases $m=2$ and $m=6$,
in which considerable simplifications occur. Their results for  $\eta_{l2}$ and $\eta_{l4}$ to order $\epsilon^2$,
agree with Sak and Grest's\cite{SG78}
but disagree with Mukamel's.\cite{Muk77}
Second, the exponent $\beta_q$ (an independent exponent
that does not follow from these correlation exponents
via a scaling law) was not considered at all by them.
Thus it is an open question whether Sak and Grest's or Mukamel's
${\mathcal O}(\epsilon^2)$ results
for $\beta_q$ with $m=2$ and $m=6$ are correct. Furthermore,
for other values of $m$,
the published ${\mathcal O}(\epsilon^2)$
results\cite{Muk77,HB78}
for the exponents
$\eta_{l2}$, $\eta_{l4}$, and $\beta_q$ remain
unchecked.

It is the purpose of this work to fill these gaps and to
determine the $\epsilon$ expansion of the critical exponents
$\eta_{l2}$, $\eta_{l4}$, and $\beta_q$
for {\em general values\/} of $m$
to  order $\epsilon^2$.

Technically, we shall employ dimensional
regularization in conjunction with minimal subtraction
of poles in $\epsilon$.
This way of fixing the counterterms
appears to us somewhat more
convenient than the use of
normalization conditions (as was done in
Refs.~\onlinecite{MC98} and \onlinecite{MC99}).
In order to overcome the rather demanding technical challenges,
we have found it useful to work directly in position space.
Thus  the Laurent expansion of the distributions to which the Feynman
graphs of the primitively divergent vertex functions correspond
in position space
must be determined to the required order in $\epsilon$.

The source of the technical difficulties is that these Feynman graphs,
at criticality,
involve  a free propagator $G(\boldsymbol{x})$
which is a {\em generalized
homogeneous} rather than a homogeneous function,
because of the {\em anisotropic\/}
scale invariance of the free theory. While such a situation is encountered
also in other cases of anisotropic scale invariance,
the scaling function associated with $G(\boldsymbol{x})$ turns out to be
a particularly complicated function
in the present case of a general $(m,d,n)$-Lifshitz point.
(For general values of $m$, it is a sum of two
generalized hypergeometric functions.)

In the next section, we recall the familiar continuum model
describing the critical behavior at a $(m,d,n)$-Lifshitz point
and discuss its renormalization. In Sec.\ \ref{sec:PertRes}
details of our calculation are presented, and
our results for the renormalization factors are derived.
Then renormalization-group equations are given in Sec.\ \ref{sec:RGE},
which are utilized to 
deduce the general scaling form of the correlation functions,
to identify the critical exponents, and to derive
their scaling laws as well as the anticipated
multi-scale-factor universality. This is followed by a presentation of our
$\epsilon$-expansion results
for the critical exponents $\eta_{l2}$, $\eta_{4l}$, and $\beta_q$.
Sec.\ \ref{sec:Concl} contains
a brief summary and concluding remarks. Finally,
there are two appendices to which some computational details
have been relegated. 

\section{The Model and its Renormalization}\label{Model}

We consider the standard continuum model representing
the universality class of a $(m,d,n)$-Lifshitz point
with the Hamiltonian
\begin{eqnarray}
{\mathcal H} [\boldsymbol{\phi}]&=&\frac{1}{2}\int d^d x\left\{
\rho_0\left({\nabla\!_\parallel}\, \boldsymbol{\phi}\right)^2
+{\sigma_0}\left({\Delta_\parallel} \boldsymbol{\phi}\right)^2
\right.\nonumber\\&&\left.\mbox{}\qquad
+\left({\nabla\!_\perp}\, \boldsymbol{\phi}\right)^2
+{\tau_0}\,\boldsymbol{\phi}^2+\frac{u_0}{12}\,|\boldsymbol{\phi}|^4\right\}\;.
\end{eqnarray}
Here $\boldsymbol{\phi}(\boldsymbol{x})=(\phi_1,\ldots,\phi_n)$
is an $n$-component order-parameter field.
The coordinate $\boldsymbol{x}\in{\mathbb{R}}^d$ has an $m$-dimensional
parallel component, $\boldsymbol{x}_\parallel$, and a ($d-m$)-dimensional
perpendicular one, $\boldsymbol{x}_\perp$. Likewise, $\nabla\!_\parallel$ and $\nabla\!_\perp$
denote the associated parallel and  perpendicular components
of the gradient operator ${\nabla}$, while $\Delta_\parallel$
means the Laplacian ${\nabla\!_\parallel}^2$.
At the level of Landau
theory, the Lifshitz point is located at $\rho_0=\tau_0=0$.

The Hamiltonian is invariant under the transformation
\begin{eqnarray}
&&\boldsymbol{x}_\parallel\to a\,\boldsymbol{x}_\parallel\;,\;\;
\boldsymbol{x}_\perp\to \boldsymbol{x}_\perp\;,\;\;
\boldsymbol{\phi}\to a^{-m/2}\,\boldsymbol{\phi}\;,\nonumber\\
&&\sigma_0\to a^4\,\sigma_0\;,\;\;
\rho_0\to a^2\,\rho_0\;,\;\;
\tau_0\to\tau_0\;,\nonumber\\
&&u_0\to a^m\,u_0\;.
\end{eqnarray}
Thus, appropriate invariant interaction constants
are $u_0\,{\sigma_0}^{-m/4}$, $\rho_0\,{\sigma_0}^{-1/2}$,
and $\tau_0$, and the dependence on the parallel
coordinates is through the invariant combination
${\sigma_0}^{-1/4}\,\boldsymbol{x}_\parallel$.

Dimensional analysis yields the dimensions $[.]$:
\begin{eqnarray}\label{diman}
[x_\parallel]&=&[\sigma_0]^{{1/ 4}}\,\mu^{-{1/ 2}}\,,\quad[x_\perp]=\mu^{-1}\,,
\nonumber \\[0em]
[\tau_0]&=&\mu^2\,,\quad [\rho_0]=[\sigma_0]^{1/2}\,\mu\;,
 \nonumber\\[0em]
[u_0]&=&[\sigma_0]^{m/4}\,\mu^\epsilon\;\mbox{ with }\;\epsilon=d^*(m)-d\;,
 \nonumber\\[0em]
[\phi_i(\boldsymbol{x})]&=&[\sigma_0]^{-{m / 8}}\,\mu^{\left(d-2-\frac{m}{2}\right)/2}\;,
\end{eqnarray}
where $\mu$ is an arbitrary momentum scale.

Let 
\begin{equation}
G_{i_1,\ldots,i_N}^{(N)}(\boldsymbol{x}_1,\ldots,\boldsymbol{x}_N)
=\langle\phi_{i_1}(\boldsymbol{x}_1)\ldots
\phi_{i_N}(\boldsymbol{x}_N)\rangle^{\mathrm cum}
\end{equation}
denote the connected $N$-point correlation functions (cumulants)
and $\Gamma_{i_1,\ldots,i_N}^{(N)}(x_1,\ldots,x_N)$
the corresponding vertex functions.
Using power counting one concludes that the ultraviolet (uv) singularities
of these functions can be absorbed through the reparameterizations
\begin{mathletters}\label{reprel}
\begin{eqnarray}
\phi&=&{Z_\phi}^{{1}/{2}}\,\phi_{\mathrm ren}\;,\\[0em]
\tau_0-\tau_{0c}&=&\mu^2\,Z_\tau\,\tau\;,\\[0em]
\sigma_0&=&Z_\sigma\,\sigma\;,\\[0em]
u_0\,{\sigma_0}^{-m/4}\,A_{d,m}&=&\mu^\epsilon\,Z_u\,u\;,
\\[0em]
\left(\rho_0-\rho_{0c}\right)\,{\sigma_0}^{-1/2}
&=&\mu\,Z_\rho\,\rho\;,
\end{eqnarray}
\end{mathletters}\noindent
where
\begin{equation}
A_{d,m}={S_{d-m}\,S_m}\,={4\,\pi^{d/2}\over
\Gamma\big({d-m\over 2}\big)\,\Gamma(m/2)}
\end{equation}
is a convenient normalization factor
we absorb in the renormalized coupling
constant. Here
\begin{equation}
S_D=\frac{2\,\pi^{D/2}}{\Gamma(D/2)}
\end{equation}
is the surface area of a $D$-dimensional unit sphere.

The quantities $\tau_{0c}$ and $\rho_{0c}$ correspond to
shifts of the Lifshitz point. In our perturbative approach based
on dimensional regularization they vanish.
If we wanted to regularize the uv singularities
via a cutoff $\Lambda$ (restricting the integrations
over parallel and perpendicular momenta by
$|{\boldsymbol{q}\/_\parallel}|\le \Lambda$ and
$|{\boldsymbol{q}\/_\perp}|\le \Lambda$),
they would be needed to absorb
uv singularities quadratic and linear in $\Lambda$,
respectively.

In the renormalization scheme we use, the renormalization
factors $Z_\phi$, $Z_\sigma$, $Z_\tau$, $Z_\rho$, and $Z_u$,
for given values of the parameters $\epsilon$, $n$, and $m$,
depend just on the dimensionless renormalized coupling
constant $u$; that is, they are independent of $\sigma$, $\tau$,
and $\rho$. This follows from the fact that the primitive
divergences of the momentum-space vertex functions
$\tilde\Gamma^{(2)}(\boldsymbol{q})$
and $\tilde\Gamma^{(4)}(\boldsymbol{q}_1,\ldots,\boldsymbol{q}_1)$,
at any order
of $u_0\,{\sigma_0}^{-m/4}$, are poles in $\epsilon$
whose residua depend
linearly on ${q_\perp}^2$, $\rho_0\,{q_\parallel}^2$,
$\sigma_0\,{q_\parallel}^4$, and $\tau_0$
in the case of the former and
are independent of these momenta and mass parameters
in the case of the latter.
Subtracting these poles minimally as usual implies
that these renormalization factors differ from $1$
through Laurent series in $\epsilon$:
\begin{eqnarray}
Z_\iota&=&1+ \sum_{p=1}^\infty {a_{\iota,p}^{(r)}(u;m,n)\, \epsilon^{-p}}
\nonumber\\
&=&1+ \sum_{r=1}^\infty\sum_{p=1}^r\,
a_{\iota,p}^{(r)}(m,n)\,\frac{u^r}{\epsilon^p}\;,\;\;
\iota=\phi,\,u,\,\tau,\,\sigma,\,\rho\,.\nonumber\\
\end{eqnarray}

\section{Outline of Computation and Perturbative Results}\label{sec:PertRes}
We shall compute the leading nontrivial contributions to these renormalization factors.
In the cases of $Z_\phi$, $Z_\sigma$, and $Z_\rho$, whose
${\mathcal O}(u)$ contributions vanish,
these are of order $u^2$; for $Z_u$ and $Z_\tau$
they are of first order in $u$.

To this end we expand about the Lifshitz point,
using the free propagator
\begin{equation}
G(\boldsymbol{x})=\int\limits_{\boldsymbol{q}}\,
\frac{{\mathrm{e}}^{i(\boldsymbol{q}_\parallel\cdot\boldsymbol{x}_\parallel+
\boldsymbol{q}_\perp\cdot\boldsymbol{x}_\perp)}}{\sigma_0\,
q_\parallel^4+q_\perp^2}\;.
\end{equation}
Here the (dimensionally regularized) momentum-space integral
is defined through
\begin{equation}
\int\limits_{\boldsymbol{q}}\ldots=
\int\limits_{\boldsymbol{q}_\parallel}\int\limits_{\boldsymbol{q}_\perp}\ldots=
\int\limits_{{\mathbb{R}}^m}\frac{d^mq}{(2\,\pi)^m}\!
\int\limits_{{\mathbb{R}}^{d-m}}\frac{d^{d-m}q}{(2\,\pi)^{d-m}}\ldots\;.
\end{equation}
Let $r_\parallel\equiv |\boldsymbol{x}_\parallel|$ and
$r_\perp\equiv |\boldsymbol{x}_\perp|$. Then
the free propagator can be written in the scaling form
\begin{eqnarray}\label{Gsf}
G(\boldsymbol{x})
&=&{r_\perp}^{-2+\epsilon}\,
\Phi\big({\sigma_0}^{-1/4}\,{r_\parallel\,{r_\perp}^{-1/2}}\big)
\end{eqnarray}
with
\begin{equation}\label{Phidef}
\Phi(\upsilon)\equiv\Phi(\upsilon;m,d)=\int\limits_{\boldsymbol{q}}\,
\frac{{\mathrm{e}}^{i(\boldsymbol{q}_\parallel\cdot\boldsymbol{\upsilon}+
\boldsymbol{q}_\perp\cdot\boldsymbol{e}_\perp)}}{q_\parallel^4+q_\perp^2}\;,
\end{equation}
where $\boldsymbol{\upsilon}\in {\mathbb R}^m$ is a vector of length $\upsilon$
and arbitrary orientation,
while $\boldsymbol{e}_\perp$ means the unit vector $\boldsymbol{x}_\perp/r_\perp$. Note that the scaling function
$\Phi$ depends parametrically on $m$ and $d$.
For the sake of brevity, we will usually suppress these variables,
writing $\Phi(\upsilon;m,d)$ only
when special values of $m$ and $d$ are chosen or when we wish
to stress the dependence on these parameters.
 
The integration over $\boldsymbol{q}_\perp$
in (\ref{Phidef}) yields
\begin{equation}\label{PhiF}
\Phi(\upsilon)=(2\,\pi)^{-{d-m\over 2}}\, \int\limits_{\boldsymbol{q}_\parallel}\,{q_\parallel}^{d-m-2}\,
{\mathrm K}_{\frac{d-m}{2}-1}({q_\parallel}^2 )\,
{\mathrm{e}}^{i\,\boldsymbol{\upsilon}\cdot\boldsymbol{q}_\parallel}\;.
\end{equation}
Upon introducing spherical coordinates
$q=|\boldsymbol{q}_\parallel|$ and $\Omega ^{(m)}=(\theta_1,\ldots,\theta_{m-1})$
for $\boldsymbol{q}_\parallel$, with $d\Omega^{(m)} =
\sin^{m-2}\theta_{m-1}\,d\theta_{m-1}\,d\Omega^{(m-1)}$,
one can perform the angular integrations. This gives
\begin{equation}\label{Phisi}
\Phi(\upsilon)={\upsilon^{-\frac{m-2}{2}}\over (2\,\pi)^{{d/2}}}\,\int\limits_0^\infty\!dq\,
q^{2-\epsilon}
\,{\mathrm J}_{m-2\over 2}(q\upsilon)\,
{\mathrm K}_{\frac{d-m}{2}-1}({q}^2 )\;.
\end{equation}
The integral remaining in (\ref{Phisi}) 
can be expressed as a combination of generalized hypergeometric
functions  ${{_{1\!}{\mathrm F}_{2}}}$ (see Appendix \ref{App:sfPhiXi}).
For special values of $m$
and $d$, the result reduces to simple expressions, which we have
gathered in Appendix \ref{App:sfPhiXi}.

The leading loop corrections to the
vertex functions $\Gamma^{(4)}$ and $\Gamma^{(2)}$
at the Lifshitz point are given in position space by the 
graphs 
\raisebox{-2.2pt}{\begin{texdraw}
\drawdim pt \setunitscale 2.5   \linewd 0.2
\move(-4 0) \move(0 3)
\move(5 0) \lellip rx:5 ry:2
\htext(-4.5 -1.2){$\boldsymbol{x}$}
\htext(11.5 -1.9){$\boldsymbol{y}$}
\move(0 0) \fcir f:1 r:0.7 \lcir r:0.7
\move(10 0) \fcir f:1 r:0.7 \lcir r:0.7
\move(16.5 0)
\end{texdraw}}
and
\raisebox{-2.2pt}{\begin{texdraw}
\drawdim pt \setunitscale 2.5   \linewd 0.2
\move(-4 0) \move(0 3)
\move(5 0) \lellip rx:5 ry:2
\move(10 0)
\rlvec(-10 0)
\htext(-4.5 -1.2){$\boldsymbol{x}$}
\htext(11.5 -1.9){$\boldsymbol{y}$}
\move(0 0) \fcir f:1 r:0.7 \lcir r:0.7
\move(10 0) \fcir f:1 r:0.7 \lcir r:0.7
\move(16.5 0)
\end{texdraw}},
which are proportional to $G^2(\boldsymbol{x}-\boldsymbol{y})$
and $G^3(\boldsymbol{x}-\boldsymbol{y})$, respectively.
Hence we must determine the Laurent expansion of these
distributions. To this end we set $\sigma_0=1$ and
consider the action $(G^s,\varphi)$ of
$G^{s}(\boldsymbol{x})$ for $s=2,3$
 on a test function $\varphi(\boldsymbol{x})$.
We substitute (\ref{Gsf}) for $G$ and use spherical coordinates
$(r_\parallel,\Omega_\parallel)$ and  $(r_\perp,\Omega_\perp)$
for the parallel and perpendicular components of $\boldsymbol{x}$,
writing $\varphi(\boldsymbol{x})=
\varphi(r_\parallel,\Omega_\parallel;r_\perp,\Omega_\perp)$.
We thus obtain
\begin{eqnarray}\label{Gsdist}
\left(G^s,\varphi\right)&=&
\left({r_\perp}^{-s(2-\epsilon)}\,
\Phi^s\big({r_\parallel\,{r_\perp}^{-1/2}}\big),\varphi\right)
\nonumber\\
&=&\int \!d^dx\,
{r_\perp}^{-s(2-\epsilon)}\,
\Phi^s\big({r_\parallel\,{r_\perp}^{-1/2}}\big)\,
\varphi(\boldsymbol{x})\nonumber\\[1em]
&=&\int\! d^{d-m} x_\perp\;
{r_\perp}^{-s(2-\epsilon)+\frac{m}{2}}\,
\psi_s(\boldsymbol{x}_\perp)\;,
\end{eqnarray}
where the functions $\psi_s(\boldsymbol{x}_\perp)
\equiv \psi_s(r_\perp,\Omega_\perp)$ are
defined through
\begin{equation}\label{psis}
\psi_s(\boldsymbol{x}_\perp)=\int\!d^m{x_\parallel}\,
\Phi^s\big(r_\parallel\big)\,
\varphi\big(
r_\parallel\,\sqrt{r_\perp},\Omega_\parallel;r_\perp,\Omega_\perp\big)\,.
\end{equation}

The final result in (\ref{Gsdist}) is the 
linear functional
$\left({r_\perp}^{-s(2-\epsilon)+\frac{m}{2}},\psi_s\right)$.
Generalized functions such as ${r_\perp}^{(\ldots)}$
and their Laurent expansions are discussed in Ref.\ \onlinecite{GS64}.
Let $\psi(\boldsymbol{x}_\perp)\equiv\psi(r_\perp,\Omega_\perp)$
be a smooth ($C^{\infty}$)
test function on ${\mathbb R}^{d-m}$ and
\begin{equation}
\overline{\psi}^{\Omega_\perp}(r_\perp)\equiv\frac{1}{S_{d-m}}\,
\int\!d\Omega_{\perp}\,\psi(r_\perp,\Omega_\perp)
\end{equation}
its spherical average. 
Then we have
\begin{eqnarray}
\left({r_\perp}^{-s(2-\epsilon)+\frac{m}{2}},\psi\right)&\equiv&
\int\!d^{d-m}{x_\perp}\,{r_\perp}^{-s(2-\epsilon)+\frac{m}{2}}\,
\psi(\boldsymbol{x}_\perp)
\nonumber \\
&=&S_{d-m}\,\int\limits_0^\infty\!dr\,
{r}^{3-2\,s+\epsilon(s-1)}\,
\overline{\psi}^{\Omega_\perp}(r)\nonumber \\
&=&
S_{d-m}\,\left({r_+}^{3-2\,s+\epsilon(s-1)},
\overline{\psi}^{\Omega_\perp}\right)\,.\nonumber\\
\end{eqnarray}
Here ${r_+}^\lambda$ is a standard generalized function in the notation
of Ref.\ \onlinecite{GS64}. Its Laurent expansion about the pole at
$\lambda=-p=-1,-2,\ldots$ reads
\begin{equation}
{r_+}^{\lambda}=\frac{(-1)^{p-1}}{(p-1)!}\,\frac{\delta^{(p-1)}(r)}{\lambda+p}
+{r_+}^{-p}+{\mathcal O}(\lambda+p)\;,
\end{equation}
where the generalized function ${r_+}^{-p}$
is defined by
\begin{eqnarray}
\left({r_+}^{-p},\varphi(r)\right)&=&\int\limits_0^\infty\!dr\,{r_+}^{-p}
\biggl[\varphi(r)-\sum_{j=0}^{p-2}\frac{r^j}{j!}\,\varphi^{(j)}(0)
\nonumber\\&&\mbox{}
-\frac{r^{p-1}}{(p-1)!}\,\varphi^{(p-1)}(0)\,\theta(1-r)
\biggr].
\end{eqnarray}

Using these results, the leading terms of the
Laurent expansions of $(G^s,\varphi)$
can be determined in a straightforward manner.
However, it should be noted that the functions
$\psi_s(\boldsymbol{x}_\perp)$ introduced in
(\ref{psis}) are not a priori guaranteed to have the usually
required strong properties of test functions
(continuous partial derivatives of all orders and
sufficiently fast decay as $|\boldsymbol{x}_\perp|\to \infty$).
In particular, one may wonder whether the dependence
on the variable $r_\parallel\,\sqrt{r_\perp}$ of $\varphi$ in
(\ref{psis}) does not imply that derivatives such as
$\nabla_\perp\psi_s$
become singular at the origin. Closer inspection reveals
that this is not the case since the problematic term
$\sim {r_\perp}^{-1}$ involves the vanishing angular integral
$\int\!d\Omega_{\parallel}\,\boldsymbol{x}_\parallel\,\varphi(\ldots)$.

One obtains
\begin{equation}\label{G2varphi}
{\left(G^2,\varphi\right)\over S_{d-m}}=
\bigg[
\frac{\psi_2(\boldsymbol{0})}{\epsilon}+
\left({r_+}^{-1},\overline{\psi_2}^{\Omega_\perp}(r)\right)
+{\mathcal O}(\epsilon)\bigg]
\end{equation}
and
\begin{equation}\label{G3varphi}
{\left(G^3,\varphi\right)\over S_{d-m}}=
\bigg[
\frac{{\overline{\psi_3}^{\Omega_\perp}}''(0)}{4\,\epsilon}+
\left({r_+}^{-3},\overline{\psi_3}^{\Omega_\perp}(r)\right)
+{\mathcal O}(\epsilon)\bigg].
\end{equation}
From its definition in (\ref{psis}) we see that the residuum
$\psi_2(\boldsymbol{0})$ on the right-hand side of (\ref{G2varphi})
reduces to a simple expression
$\propto \varphi(\boldsymbol{0})$.
We thus arrive at the expansion
\begin{equation}\label{cu}
{G^2(\boldsymbol{x})\over A_{d,m} }
=\frac{J_{0,2}(m,d^*)}{\epsilon}\,\delta(\boldsymbol{x})
\,
+{\mathcal O}(\epsilon^0)\;,
\end{equation}
where $J_{0,2}$ is a particular one of the integrals
\begin{equation}\label{jps}
J_{p,s}(m,d)\equiv\int\limits_0^\infty \!\upsilon^{m-1+p}\,\Phi^s(\upsilon;m,d)\,d\upsilon\;.
\end{equation}

In order to convert the Laurent expansion (\ref{G3varphi})
into one for $G^3(\boldsymbol{x})$,
we must compute ${\overline{\psi_3}^{\Omega_\perp}}''(0)$.
This in turn requires the calculation of the following angular average:
\begin{eqnarray}
\lefteqn
{\left.{\partial^2\over\partial r^2}\,\overline{%
\varphi\big(r_\parallel\sqrt{r},\Omega_\parallel;
r,\Omega_\perp\big)}^{\Omega_\parallel,\Omega_\perp}\right|_{r=0}}
&&
\nonumber\\[0.5em]
&=&\frac{2}{4!}\,\overline{\left(%
\boldsymbol{x}_\parallel\cdot\nabla_\parallel\right)^4\,
\varphi(\boldsymbol{0})}^{\Omega_\parallel,\Omega_\perp}
+ \overline{\left(\boldsymbol{e}_\perp\cdot\nabla_\perp\right)^2\,
\varphi(\boldsymbol{0})}^{\Omega_\parallel,\Omega_\perp}
\nonumber\\[0.5em]
&=&\frac{{r_\parallel}^4\,
({\triangle_\parallel}^2\varphi)(\boldsymbol{0})}{4m(m+2)}
+\frac{(\triangle_\perp\varphi)(\boldsymbol{0})}{d-m}\;.
\end{eqnarray}
Using this in conjunction with (\ref{G3varphi})
gives
\begin{eqnarray}\label{cfs}
{G^3(\boldsymbol{x})\over A_{d,m}}
&=&{J_{4,3}(m,d^*)\,{\triangle_\parallel}^2\delta(\boldsymbol{x})\over 16\,m(m+2)\,\epsilon}
\nonumber\\&&\mbox{}
+{J_{0,3}(m,d^*)\,{\triangle_\perp}\delta(\boldsymbol{x})%
\over 4\,(d^*-m)\,\epsilon}
+{\mathcal O}(\epsilon^0)\,.
\end{eqnarray}

In order to compute the ${\mathcal O}(u^2)$ term of $Z_\rho$,
we consider the two-point vertex function with an insertion
of the operator $\frac{1}{2}\int d^dx ({\nabla_{\|}}\phi)^2$ (to which
$\rho_0$ couples). We represent such an insertion
by the vertex 
\raisebox{-1pt}{\begin{texdraw}
\drawdim pt \setunitscale 2.5   \linewd 0.2
\move(5 2)
\fcir f:0 r:0.7
\move(3.85 1)\rlvec(0 2)
\move(6.15 1)\rlvec(0 2)
\move(3 2) \rlvec(4 0)
\move(8 2)
\end{texdraw}}.
At the Lifshitz point $\tau=\rho=0$, the leading
nontrivial contribution to this vertex function is given by the
 two-loop graph
\raisebox{-2.2pt}{\begin{texdraw}
\drawdim pt \setunitscale 2.5   \linewd 0.2
\move(-5 0) \move(0 3)
\move(5 0) \lellip rx:5 ry:2
\move(5 2)
\fcir f:0 r:0.7
\move(3.85 1)\rlvec(0 2)
\move(6.15 1)\rlvec(0 2)
\move(10 0)
\rlvec(-10 0)
\move(10 0) \fcir f:1 r:0.7 \lcir r:0.7
\move(0 0) \fcir f:1 r:0.7 \lcir r:0.7
\htext(-4.5 -1.2){$\boldsymbol{x}$}
\htext(11.5 -1.2){$\boldsymbol{0}$}
\move(15.5 0)
\end{texdraw}}. The upper line involves the
convolution
\begin{equation}\label{Xiintro}
-(\nabla_\| G*{\nabla_\|}G)(\boldsymbol{x})={r_\perp}^{-1+\epsilon}\,
\Xi\big({\sigma_0}^{-1/4}\,{r_\parallel\,{r_\perp}^{-1/2}}\big)\,,
\end{equation}
where
\begin{equation}\label{Xidef}
\Xi(\upsilon)\equiv\Xi(\upsilon;m,d)= \int\limits_{\boldsymbol{q}}\,
\frac{{q_\|}^2\,
{\mathrm{e}}^{i(\boldsymbol{q}_\parallel\cdot\boldsymbol{\upsilon}+
\boldsymbol{q}_\perp\cdot\boldsymbol{e}_\perp)}}
{\left({q_\parallel}^4+{q_\perp}^2\right)^2}
\end{equation}
is the analog of the scaling function $\Phi(\upsilon)$ (cf.\ (\ref{Gsf})).
Proceeding as in the case of the latter, one obtains
\begin{eqnarray}\label{sfXi}
\Xi(\upsilon)&=&
\frac{1}{2\,(2\pi)^{{d-m\over 2}}}
\int\limits_{\boldsymbol{q}_\|}\!
{q_\parallel}^{d-m-2}\,
{\mathrm K}_{\frac{d-m-4}{2}}({q_\parallel}^2 )\,
{\mathrm{e}}^{i\,\boldsymbol{\upsilon}\cdot\boldsymbol{q}_\parallel}\nonumber\\
&=&{\upsilon^{-\frac{m-2}{2}}\over 2\, (2\,\pi)^{{d/2}}}\,\int\limits_0^\infty\!dq\,
q^{2-\epsilon}
\,{\mathrm J}_{m-2\over 2}(q\upsilon)\,
{\mathrm K}_{\frac{d-m-4}{2}}({q}^2 )\;.\nonumber\\
\end{eqnarray}
The remaining single integral can again be expressed
in terms of generalized hypergeometric functions.
The corresponding general expression, as well as the
simpler ones to which this reduces for special
values of $m$ and $d$,  may be found in Appendix \ref{App:sfPhiXi}).

The required graph
\raisebox{-2.2pt}{\begin{texdraw}
\drawdim pt \setunitscale 2.5   \linewd 0.2
\move(-5 0) \move(0 3)
\move(5 0) \lellip rx:5 ry:2
\move(5 2)
\fcir f:0 r:0.7
\move(3.85 1)\rlvec(0 2)
\move(6.15 1)\rlvec(0 2)
\move(10 0)
\rlvec(-10 0)
\move(10 0) \fcir f:1 r:0.7 \lcir r:0.7
\move(0 0) \fcir f:1 r:0.7 \lcir r:0.7
\htext(-4.5 -1.2){$\boldsymbol{x}$}
\htext(11.5 -1.2){$\boldsymbol{0}$}
\move(15.5 0)
\end{texdraw}}
is proportional to the distribution
\begin{equation}\label{LexpD}
D(\boldsymbol{x})=-
G^2(\boldsymbol{x})\,({\nabla_\|}G*{\nabla_\|}G)(\boldsymbol{x})\;.
\end{equation}
whose pole term can be worked out in a straightforward fashion
by the techniques employed above.
One finds
\begin{equation}\label{cr}
\frac{-G^2(\boldsymbol{x})\,({\nabla_\|}G*{\nabla_\|}G)(\boldsymbol{x})}{A_{d,m}}=
{I_1(m,d^*)\,{\triangle_\parallel}%
\delta(\boldsymbol{x})\over 4\,m\,\epsilon}
+{\mathcal O}(\epsilon^0)
\end{equation}
with
\begin{equation}\label{i1}
I_1(m,d)\equiv\int\limits_0^\infty\!\upsilon^{m+1}\,
\Phi^2(\upsilon;m,d)\,\Xi(\upsilon;m,d)\,d\upsilon\;.
\end{equation}

A convenient way of computing the renormalization factor
$Z_\tau$ is to consider the
vertex function $\Gamma^{(2,1)}$ with a single 
insertion of the operator $\frac{1}{2}\,\phi(\boldsymbol{y})^2$,
which we depict as
\raisebox{-1pt}{\begin{texdraw}
\drawdim pt \setunitscale 2.5   \linewd 0.2
\move(5 2)
\fcir f:0 r:0.7
\move(3.5 2) \rlvec(3 0) 
\rlvec(-1.5 0) \lpatt(0.5 0.5)
\rlvec(0 2.5)
\htext(3.6 5){$\small\boldsymbol{y}$}
\move(7.5 2) \move(3.6 7)
\end{texdraw}}.
Its one-loop contribution
\raisebox{-12pt}{\begin{texdraw}
\drawdim pt \setunitscale 2.5   \linewd 0.2
\move(-5 0)
\move(0 2) \lellip rx:4 ry:2
\move(0 0) \fcir f:1 r:0.7 \lcir r:0.7
\move(0 4) \fcir f:0 r:0.7
\lpatt(0.5 0.5)
\rlvec(0 2.5)
\htext(-3 4.4){$\small\boldsymbol{y}$}
\htext(-1.35 -2.9){$\small\boldsymbol{x}$}
\move(5 0)\move(0 -3.5)
\end{texdraw}}
is proportional to
$G^2(\boldsymbol{x}-\boldsymbol{y})$. Hence its
Laurent expansion follows from that of the latter quantity.

Let us introduce coefficients $b_\iota(m)$ for the leading nontrivial contributions to the renormalization factors $Z_\iota$,
writing these in the form
\begin{equation}\label{Zu}
Z_u=1+b_u(m)\,\frac{n{+}8}{9}\,\frac{u}{\epsilon}+{\mathcal O}(u^2)\;,
\end{equation}
\begin{equation}\label{Ztau}
Z_\tau=1+b_\tau(m)\,\frac{n{+}2}{3}\,\frac{u}{\epsilon}+{\mathcal O}(u^2)\;,
\end{equation}
and
\begin{equation}\label{Zvarsigma}
Z_{\varsigma}=1+b_{\varsigma}(m)\,\frac{n{+}2}{3}\,\frac{u^2}{\epsilon}
+{\mathcal O}(u^3)\;,
\quad\varsigma=\phi,\sigma,\rho\,.
\end{equation}

From the pole terms of $G^2(\boldsymbol{x}-\boldsymbol{y})$
given in (\ref{cu}) one easily deduces that
\begin{equation}\label{bubtau}
b_u(m)=3\,b_\tau(m)=\frac{3}{2}\,J_{0,2}(m,d^*)\;.
\end{equation}
The pole terms proportional to ${\triangle_\perp}\delta(\boldsymbol{x})$,
${\triangle_\parallel}^2\,\delta(\boldsymbol{x})$,
and ${\triangle_\parallel}\,\delta(\boldsymbol{x})$ of the
two-loop graphs considered above
are absorbed by counterterms involving
the renormalization factors $Z_\phi$,
$\check{Z}_\sigma\equiv Z_\sigma\,Z_\phi$,
and $\check{Z}_\rho\equiv Z_\rho\,Z_\phi\,{Z_\sigma}^{1/2}$,
respectively.
Utilizing the Laurent expansions (\ref{cfs}) and (\ref{LexpD}),
one finds that their coefficients are given by
\begin{equation}\label{bphi}
b_\phi(m)=-{1\over 24}\,{1\over d^*{-}m}\,{J_{0,3}(m,d^*)\over A_{d^*,m}}\;,
\end{equation}
\begin{equation}\label{cbsigma}
\check{b}_\sigma(m)={1\over 96}\,{1\over m(m{+}2)} \,\frac{J_{4,3}(m,d^*)}{A_{d^*,m}}\;,
\end{equation}
and
\begin{equation}\label{cbrho}
\check{b}_\rho(m)={1\over 8m}\,\frac{I_1(m,d^*)}{A_{d^*,m}}\;.
\end{equation}
The coefficients $b_\sigma$ and $b_\rho$ are related to these via
\begin{equation}
b_\sigma(m)=\check{b}_\sigma(m)-b_\phi(m)
\end{equation}
and
\begin{equation}
 b_\rho(m)=\check{b}_\rho(m)-{1\over 2}\,b_\phi(m)
-{1\over 2}\,\check{b}_\sigma(m)\;.
\end{equation}

\section{Renormalization Group Equations and
$\boldsymbol{\epsilon}$-Expansion Results}\label{sec:RGE}

The reparameterizations  (\ref{reprel})
yield the following relations between bare and
 renormalized correlation and vertex functions
\begin{mathletters} \label{RGE}
\begin{eqnarray}\label{RGEG}
G^{(N)}(\boldsymbol{x}_\parallel,\boldsymbol{x}_\perp)&=&
{Z_\phi}^{N/2}\,
G_{\mathrm {ren}}^{(N)}(\boldsymbol{x}_\parallel,\boldsymbol{x}_\perp)
\;,\\[1em]
\Gamma^{(N)}(x_\parallel,x_\perp)&=&
{Z_\phi}^{-N/2}\,\Gamma_{\mathrm {ren}}^{(N)}(x_\parallel,
x_\perp)\;,
\end{eqnarray}
\end{mathletters}\noindent%
where  $\boldsymbol{x}_\parallel$ and $\boldsymbol{x}_\perp$
stand for the set of all parallel and perpendicular coordinates on
which  $G^{(N)}$ and $\Gamma^{(N)}$ depend.
For conciseness, we have suppressed the tensorial indices
$i_1,\ldots,i_N$ of these functions and will generally do so
below.

Upon exploiting the invariance of the bare functions under a change
 $\mu\to\bar{\mu}(\ell)=\mu\,\ell$ of the momentum scale
in the usual fashion, one arrives at 
the renormalization-group equations
\begin{eqnarray}
\left[{\mathcal{D}}_\mu+\frac{N}{2}\,\eta_\phi\right]G_{\mathrm ren}^{(N)}&=&0\;,
\\[1em]
\left[{\mathcal{D}}_\mu-\frac{N}{2}\,\eta_\phi\right]
\Gamma_{\mathrm ren}^{(N)}&=&0
\end{eqnarray}
with
\begin{eqnarray}
{\mathcal{D}}_\mu&=&\mu\,\partial_\mu +\beta_u\,\partial_u 
-\eta_\sigma\,\sigma\,\partial_\sigma
\nonumber\\ \mbox{}&&
-(2+\eta_\tau)\,\tau\,\partial_\tau
-(1+\eta_\rho)\,\rho\,\partial_\rho\;,
\end{eqnarray}
where the beta and eta functions are defined by
\begin{eqnarray}\label{betadef}
\beta_u&\equiv&\left.\mu\,\partial_\mu\right|_0 u=-u\,\left[\epsilon+\eta_u(u)\right]
\end{eqnarray}
and
\begin{eqnarray}\label{etadef}
\eta_\iota\equiv\left.\mu\,\partial_\mu\right|_0\ln Z_\iota\;,
\quad\iota=\phi,\,\sigma,\,\rho,\,\tau,\,u\;,
\end{eqnarray}
respectively. Here
 $\left.\partial_\mu\right|_0$ means a derivative at fixed
bare variables
$\mu_0$, $\rho_0$, $\sigma_0$, and $\tau_0$.
Owing to our use of the minimal subtraction procedure,
the functions $\eta_\iota$ can be expressed
in terms of the residua $a_{\iota,1} (u;m,n)$ as
\begin{equation}\label{etares}
\eta_\iota (u) = -u\,\frac{da_{\iota,1}}{du} \;,
 \quad\iota=\phi,\,\sigma,\,\rho,\,\tau,\,u.
\end{equation}

To solve the renormalization-group equations (\ref{RGE}) via
characteristics, we introduce flowing variables through
\begin{equation}\label{eq:uflow}
\ell\,\frac{d}{d\ell}\,\bar{u}(\ell)=\beta_u[\bar{u}(\ell)]\;,\quad \bar{u}(1)=u\;,
\end{equation}
\begin{equation}\label{eq:sigmaflow}
\ell\,\frac{d}{d\ell}\,\bar{\sigma}(\ell)=-\eta_\sigma(\bar{u})\,\sigma\;,
\quad \bar{\sigma}(1)=\sigma\;,
\end{equation}
\begin{equation}\label{eq:rhoflow}
\ell\,\frac{d}{d\ell}\,\bar{\rho}(\ell)=
-\left[1+\eta_\rho(\bar{u})\right]\rho\;,
\quad \bar{\rho}(1)=\rho\;,
\end{equation}
and
\begin{equation}\label{eq:tauflow}
\ell\,\frac{d}{d\ell}\,\bar{\tau}(\ell)=
-\left[2+\eta_\tau(\bar{u})\right]\tau\;,
\quad \bar{\tau}(1)=\tau\;.
\end{equation}

The flow equation (\ref{eq:uflow})
for the running coupling constant $\bar{u}(\ell)$
can be solved for $\ell$ to obtain
\begin{equation}
\ln\ell = \int\limits_u^{\bar{u}}\!{dx\over \beta_u(x)}\;.
\end{equation}
For $\epsilon>0$, the beta function $\beta_u(u)$ is
known to have a nontrivial zero $u^*$, corresponding
to an infrared-stable fixed point. Expanding
about this fixed point gives the familiar
asymptotic form
\begin{equation}
\bar{u}(\ell)\mathop{=}\limits_{\ell\to 0}u^*+(u-u^*)\,{\ell}^{\omega_u}+{\mathcal O}(\ell^{2\,\omega_u})
\end{equation}
in the infrared limit $\ell\to 0$,
where
\begin{equation}
\omega_u\equiv \frac{d\beta_u}{du}(u^*)
\end{equation}
is positive.

The solutions to the other flow equations,
(\ref{eq:sigmaflow})--(\ref{eq:tauflow}), can be conveniently
written in terms of the anomalous dimensions
$\eta_\iota^*\equiv\eta_\iota(u^*)$ and the
renormalization-group-trajectory integrals
\begin{equation}
E_{\iota}[\bar{u},u]=\exp\left\{
\int\limits_u^{\bar{u}(\ell)}\!{dx}\,
{\eta_\iota^*-\eta_\iota(x)\over \beta_u(x)}
\right\}\;,\;\;\iota=\phi,\,\sigma,\,\rho,\,\tau\,,
\end{equation}
which approach nonuniversal constants
\begin{equation}
E_\iota^*(u)\equiv E_{\iota}(u^*,u)
\;,\;\;\iota=\phi,\,\sigma,\,\rho,\,\tau\,,
\end{equation}
in the infrared limit $\ell\to 0$.

One finds
\begin{eqnarray}
\bar{\sigma}(\ell)&=&\ell^{-\eta_\sigma^*}\,
E_\sigma[\bar{u}(\ell),u]\,\sigma\nonumber\\
&\mathop{\approx}\limits_{\ell\to 0}&
\ell^{-\eta_\sigma^*}\,E_\sigma^*(u)\,\sigma\;,
\end{eqnarray}
\begin{eqnarray}
\bar{\rho}(\ell)&=&\ell^{-(1+\eta_\rho^*)}\,
E_\rho[\bar{u}(\ell),u]\,\rho\nonumber\\
&\mathop{\approx}\limits_{\ell\to 0}&
\ell^{-(1+\eta_\rho^*)}\,E_\rho^*(u)\,\rho\;,
\end{eqnarray}
and
\begin{eqnarray}
\bar{\tau}(\ell)&=&\ell^{-(2+\eta_\tau^*)}\,
E_\tau[\bar{u}(\ell),u]\,\tau \nonumber\\
&\mathop{\approx}\limits_{\ell\to 0}&
\ell^{-(2+\eta_\tau^*)}\,E_\tau^*(u)\,\tau\;.
\end{eqnarray} 
\end{multicols}
Solving the RG equation (\ref{RGEG}) in terms of characteristics
yields
\begin{eqnarray}\label{eq:Gflowsol}
&&G^{(N)}_{\mathrm ren}(\boldsymbol{x}_\parallel,\boldsymbol{x}_\perp;%
\rho,\tau,u,\sigma,\mu)
=\left[{\ell^{\eta_\phi^*}\over E_\phi(\bar{u},u)}\right]^{N/2}
G^{(N)}_{\mathrm ren}(\boldsymbol{x}_\parallel,\boldsymbol{x}_\perp;%
\bar{\rho},\bar{\tau},\bar{u},\bar{\sigma},\mu\,\ell)\nonumber\\
&&\quad =\left[{(\mu\,\ell)^{d-2-{m\over 2}}\,\ell^{\eta_\phi^*}\over \bar{\sigma}^{m/4}\,E_\phi(\bar{u},u)}\right]^{N/2}
G^{(N)}_{\mathrm ren}\!\left[
\left({\mu^2\,\ell^2/{\bar{\sigma}}}
\right)^{1/4}\,\boldsymbol{x}_\parallel,\mu\,\ell\,\boldsymbol{x}_\perp;%
\bar{\rho},\bar{\tau},\bar{u},1,1\right]\;.
\end{eqnarray}
To obtain the second equality, we have used the relation
\begin{eqnarray}
G^{(N)}_{\mathrm ren}(\boldsymbol{x}_\parallel,\boldsymbol{x}_\perp;%
\bar{\rho},\bar{\tau},\bar{u},\bar{\sigma},\bar{\mu})
&=&\left[\bar{\mu}^{d-2-\frac{m}{2}}\,\bar{\sigma}^{-m/4}\right]^{N/2}\,
G^{(N)}_{\mathrm ren}
(\bar{\sigma}^{-1/4}\,\bar{\mu}^{1/2}\,\boldsymbol{x}_\parallel,\bar{\mu}\,\boldsymbol{x}_\perp;%
\bar{\rho},\bar{\tau},\bar{u},1,1)\;,
\end{eqnarray}
implied by our dimensional considerations (\ref{diman}).

Let us assume that the function
$G^{(N)}_{\mathrm ren}$
on the right-hand side of (\ref{eq:Gflowsol})
has a nonvanishing limit $\bar{u}\to u^*$
for $\epsilon>0$. This assumption is in conformity
with, and can be checked by, RG-improved perturbation theory.
We choose $\ell=\ell_\tau$ such that
\begin{equation}
\bar{\tau}(\ell_\tau)=\pm 1 \quad\mbox{ for } \pm\tau >0
\end{equation}
and consider the limit $\tau\to 0\pm$.
To write the resulting asymptotic form of $G^{(N)}_{\mathrm ren}$
in a compact fashion, we introduce
the correlation-length exponents
\begin{equation}\label{nul2}
\nu_{l2}=\frac{1}{2+\eta_\tau^*}
\end{equation}
and
\begin{equation}\label{nul4}
\nu_{l4}=\frac{2+\eta_\sigma^*}{4(2+\eta_\tau^*)}\;,
\end{equation}
the crossover exponent
\begin{equation}\label{coexp}
\varphi=\nu_{l2}\,(1+\eta_\rho^*)\;,
\end{equation}
as well as the correlation lengths
\begin{equation}
\xi_\perp\equiv \mu^{-1}\,\ell_\tau\approx \mu^{-1}\,\left[E_\tau^*(u)\,|\tau|\right]^{-\nu_{l2}}
\end{equation}
and
\begin{equation}
\xi_\parallel\equiv \left[{\bar{\sigma}(\ell_\tau)\over \mu^2\,{\ell_\tau}^2}\right]^{1/4}\approx \mu^{-1/2}\,\left[E_\sigma^*(u)\,\sigma\right]^{1/4}\,
\left[E_\tau^*(u)\,|\tau|\right]^{-\nu_{l4}}\;.
\end{equation}

In terms of these quantities the asymptotic critical behavior
of $G^{(N)}_{\mathrm ren}$ becomes
\begin{equation}\label{Gscf}
G^{(N)}_{\mathrm ren}(\boldsymbol{x}_\parallel,\boldsymbol{x}_\perp;\rho,\tau,u,\sigma,\mu)\approx
\left[{\mu^{-\eta_\phi^*}\over {E_\phi^*}}
\,{\xi_\perp}^{-(d-m-2+\eta_\phi^*)}\,
{\xi_\parallel}^{-m}\right]^{N/2}
{\mathcal G}_\pm^{(N)}\!\left[
{\boldsymbol{x}_\parallel\over \xi_\parallel},{\boldsymbol{x}_\perp\over \xi_\perp};E_\rho^*\,\rho\,(\mu\,\xi_\perp)^{\varphi/\nu_{l2}}\right]
\end{equation}
with
\begin{equation}\label{scfunc}
{\mathcal G}_\pm^{(N)}(\boldsymbol{x}_\parallel,\boldsymbol{x}_\perp;\rho)\equiv
G^{(N)}_{\mathrm ren}(\boldsymbol{x}_\parallel,\boldsymbol{x}_\perp;\rho,\pm 1,u^*,1,1)\;.
\end{equation}
\begin{multicols}{2}
The result is the scaling form expected according to
the phenomenological theory of scaling. As it shows, the scaling function
${\mathcal G}_\pm^{(N)}$ is universal, up to a redefinition of the nonuniversal
metric factors associated with the relevant scaling fields,
i.e., $E_\sigma ^*$, $E_\rho ^*$, $E_\tau ^*$, and $E_\phi ^*$.
(Note that $E_\phi ^*$, whose change would affect the overall amplitude
of ${\mathcal G}_\pm^{(N)}$, as usual corresponds to a metric factor associated
with the magnetic scaling field; see, e.g., Ref.\ \onlinecite{Die86a}.)

The correlation exponents $\eta_{l2}$ and $\eta_{l4}$ are given by
\begin{equation}\label{etal2}
\eta_{l2}=\eta_\phi^*
\end{equation}
and
\begin{equation}\label{etal4}
\eta_{l4}=4\,\frac{\eta_\sigma^*+\eta_\phi^*}{2+\eta_\sigma^*}\;.
\end{equation}
This can be seen either by taking the Fourier transform of
the above result (\ref{Gscf}) with $N=2$ or else by solving directly
the renormalization-group equation of
$\tilde{\Gamma}^{(2)}_{\mathrm ren}
({\boldsymbol{q}}_\parallel,{\boldsymbol{q}}_\perp)$.
In order to identify the wave-vector exponent $\beta_q$,
we utilize the scaling form
\begin{equation}
\tilde{\Gamma}^{(2)}_{\mathrm ren}
({\boldsymbol{q}}_\parallel,{\boldsymbol{q}}_\perp;
\tau,\rho,u)\approx |\tau|^{\gamma}\;
\Upsilon_\pm(q_\parallel\xi_\|,q_\perp\xi_\perp;
\rho\,|\tau|^{-\varphi})
\end{equation}
of the inverse susceptibility $\tilde{\Gamma}^{(2)}$
and argue as in Ref.\ \onlinecite{Muk77}:
On the helical branch $T_{\text{hel}}(\rho)$ of the critical
line, the inverse susceptibility vanishes at $\boldsymbol{q}_c=(\boldsymbol{q}^c_{\|},\boldsymbol{0})\ne \boldsymbol{0}$.
Hence in the scaling regime, the line $T_{\text{hel}}(\rho)$ is
determined by the zeroes of the
scaling function
$\Upsilon(p,0,\varrho)$. Denoting these as $p_c$ and $\varrho_c$,
we obtain the relations
\begin{equation}
q^c_{\|}= {p_c\, {\xi_\|}^{-1}}\sim p_c|\tau|^{\nu_{l4}}
\end{equation}
and
\begin{equation}
\rho=\varrho_c|\tau|^{\varphi}\;,
\end{equation}
which yield
\begin{equation}
q_\|^c\sim |\tau|^{\beta_q}
\end{equation}
with
\begin{equation}\label{betaq}
\beta_q=\frac{\nu_{l4}}{\varphi}=\frac{2+\eta_\sigma^*}{4(1+\eta_\rho^*)}\;,
\end{equation}
where the last equality follows upon substitution of (\ref{coexp})
and (\ref{nul4}) for $\varphi$ and $\nu_{l4}$, respectively.

To compute the exponent functions (\ref{etadef})
and the beta function (\ref{betadef}), we
insert the residua of the
renormalization factors (\ref{Zu})--(\ref{Zvarsigma})
into (\ref{etares}) and
express $b_\tau$ in terms of $b_u$ using (\ref{bubtau}).
We thus obtain
\begin{equation}\label{etavarsigma}
\eta_\varsigma(u) = -2\,{n{+}2\over 3}\,b_\varsigma(m)\,u^2+{\mathcal O}(u^3)\,,
\quad \varsigma=\phi,\sigma,\rho\,,
\end{equation}
\begin{equation}\label{etatau}
\eta_\tau(u)=- {1\over 3}\,\frac{n{+}2}{3}\,b_u (m)\,u+{\mathcal O}(u^2)\;,
\end{equation}
and
\begin{equation}\label{betau}
\beta_u(u)=-u\left[\epsilon- \frac{n{+}8}{9}\;b_u(m)\,u+{\mathcal O}(u^2)\right].
\end{equation}

From the last equation we can read off the $\epsilon$ expansion of
$u^*$, the nontrivial zero of $\beta_u$:
\begin{equation}\label{ustar}
u^*={9\over n{+}8}\, {\epsilon\over b_u(m)}+{\mathcal O}\big(\epsilon^2\big)\,.
\end{equation}

Evaluation of the above exponent functions at this fixed-point value
 gives us
the $\epsilon$ expansions of the anomalous
dimensions $\eta_\iota^*$. 
Substituting  these into the expressions (\ref{nul2})--(\ref{coexp}),
 (\ref{etal2}), (\ref{etal4}), and  (\ref{betaq}) for the critical  exponents
yields
\begin{equation}\label{nul2exp}
\nu_{l2}={1\over 2}+{n{+}2\over 4(n{+}8)}\,\epsilon+{\mathcal O}(\epsilon^2)\;,
\end{equation}
\begin{eqnarray}\label{nul4exp}
\frac{\nu_{l4}}{\nu_{l2}}&=&\frac{1}{2}+{27(n{+}2)\over (n{+}8)^2}\;
{b_{\phi}(m){-}\check{b}_{\sigma}(m)\over {2\,b_u(m)}^2}\;\epsilon^2
+{\mathcal O}(\epsilon^3)
\nonumber\\&=&
\frac{1}{2}+{\mathcal O}(\epsilon^3)
-
{27(n{+}2)\over  (n{+}8)^2}\,{\epsilon^2\over 4}
\cases{0.02152&for $m=1$,\cr
0.02195&for $m=2$,\cr
0.02231&for $m=3$,\cr
0.02263&for $m=4$,\cr
0.02290&for $m=5$,\cr
0.02313&for $m=6$,\cr
}
\nonumber\\
\end{eqnarray}
\begin{eqnarray}\label{etal2exp}
\eta_{l2}&=&-2\,{27(n{+}2)\over (n{+}8)^2}\;
{b_\phi(m)\over {b_u(m)}^2}\;\epsilon^2
+{\mathcal O}(\epsilon^3)\nonumber\\
&=&{\mathcal O}(\epsilon^3)+{27(n{+}2)\over (n{+}8)^2}\;\epsilon^2
\cases{0.01739&for $m=1$,\cr
0.01646&for $m=2$,\cr
0.01564 &for $m=3$,\cr
0.01488&for $m=4$,\cr
0.01418&for $m=5$,\cr
0.01353&for $m=6$,\cr
}
\nonumber\\
\end{eqnarray}
\begin{eqnarray}\label{etal4exp}
\eta_{l4}&=&-4\,{27(n{+}2)\over (n{+}8)^2}\;{\check b_\sigma(m)\over {b_u(m)}^2}\;\epsilon^2
+{\mathcal O}(\epsilon^3)
\nonumber\\
&=&{\mathcal O}(\epsilon^3)-{27(n{+}2)\over (n{+}8)^2}\;\epsilon^2
\cases{ 0.00827&for $m=1$,\cr
0.01097 &for $m=2$,\cr
0.01334 &for $m=3$,\cr
0.01548&for $m=4$,\cr
0.01743&for $m=5$,\cr
0.01920&for $m=6$,\cr
}
\nonumber\\
\end{eqnarray}
\begin{eqnarray}\label{varphiexp}
\frac{\varphi}{\nu_{l2}}&=&1+\,{27(n{+}2)\over(n{+}8)^2}\;
{b_\phi(m){-}2 \check{b}_\rho(m){+}\check{b}_\sigma(m)\over {b_u(m)}^2}\;\epsilon^2
+{\mathcal O}(\epsilon^3)
\nonumber\\
&=&1+{\mathcal O}(\epsilon^3)-{27(n{+}2)\over (n{+}8)^2}\;\epsilon^2
\cases{ 0.02781&for $m=1$,\cr
0.05487 &for $m=2$,\cr
0.07856 &for $m=3$,\cr
0.09980&for $m=4$,\cr
0.11904&for $m=5$,\cr
0.13658&for $m=6$,\cr
}
\nonumber\\
\end{eqnarray}
and
\begin{eqnarray}\label{betaqexp}
\beta_q&=&{1\over 2}+{27(n{+}2)\over(n{+}8)^2}\;
{\check b_\rho(m){-}\check b_\sigma(m)\over {b_u(m)}^2}\;\epsilon^2
+{\mathcal O}(\epsilon^3)\nonumber\\
&=&{1\over 2}+{\mathcal O}(\epsilon^3)+{27(n{+}2)\over(n{+}8)^2}\;
\epsilon^2
\cases{ 0.00852&for $m=1$.\cr
0.02195 &for $m=2$.\cr
0.03370 &for $m=3$.\cr
0.04424&for $m=4$.\cr
0.05379&for $m=5$.\cr
0.06251&for $m=6$.\cr
}
\nonumber\\
\end{eqnarray}

We have expressed the results in terms of the coefficients $b_u(m)$,
$b_\phi(m)$, $\check{b}_\sigma(m)$, and $\check{b}_\rho(m)$,
which according to (\ref{bubtau})--(\ref{cbrho}) are
proportional to the integrals
$J_{0,2}(m,d^*)$, $J_{0,3}(m,d^*)$, $J_{4,3}(m,d^*)$,
and $I_1(m,d^*)$, respectively. 
These integrals are defined by (\ref{jps}) and (\ref{i1}).
The first one of them---the one-loop integral $J_{0,2}(m,d)$---%
is analytically computable\cite{J02calc}
for general values of $d$ and $m$.
The result is
\begin{equation}\label{J02}
J_{0,2}(m,d)={2^{-2-\epsilon}\over (2\pi)^d}\,
\Gamma^2\Big(1{-}{\epsilon\over 2}\Big)\,
{\Gamma(2{-}{m\over 4}{-}\epsilon)\over\Gamma(2{-}\epsilon)}
\,
\Gamma\Big({m\over 4}\Big)\,,
\end{equation}
giving
\begin{equation}
b_u(m)=\frac{3}{8}\,
\frac{\Gamma\left(2 {-}\frac{m}{4}\right)\,
  \Gamma\left(\frac{m}{4}\right)}{(2\pi)^{4+\frac{m}{2}}}\,.
\end{equation}
The fixed-point value that results when this value of $b_u(m)$ is
inserted into (\ref{ustar}) is
consistent with the one found in calculations based on
Wilson's momentum-shell integration method.\cite{Muk77}

The integrals $J_{0,3}(m,d^*)$, $J_{4,3}(m,d^*)$, and $I_1(m,d^*)$,
and hence the coefficients $b_\phi(m)$, $\check{b}_\sigma(m)$, and
$\check{b}_\rho(m)$,
can be calculated numerically for any desired value of $m$, using the
explicit expressions for the scaling functions $\Phi(\upsilon;m,d^*)$ and
$\Xi(\upsilon;m,d^*)$ given in (\ref{Phid*}) and (\ref{Xid*})
of Appendix \ref{App:sfPhiXi}. (As discussed there, the numerical evaluation of
these integrals for general values of $m$ requires some care because
$\Phi(\upsilon;m,d^*)$ is a difference of two terms, each of which
grows exponentially as $\upsilon\to\infty$.)
In this manner one arrives at the values of the $\epsilon^2$ terms
given in the second lines of  (\ref{nul4exp})--(\ref{betaqexp}).

\begin{figure}[htb] 
\begin{center}
\epsfig{file=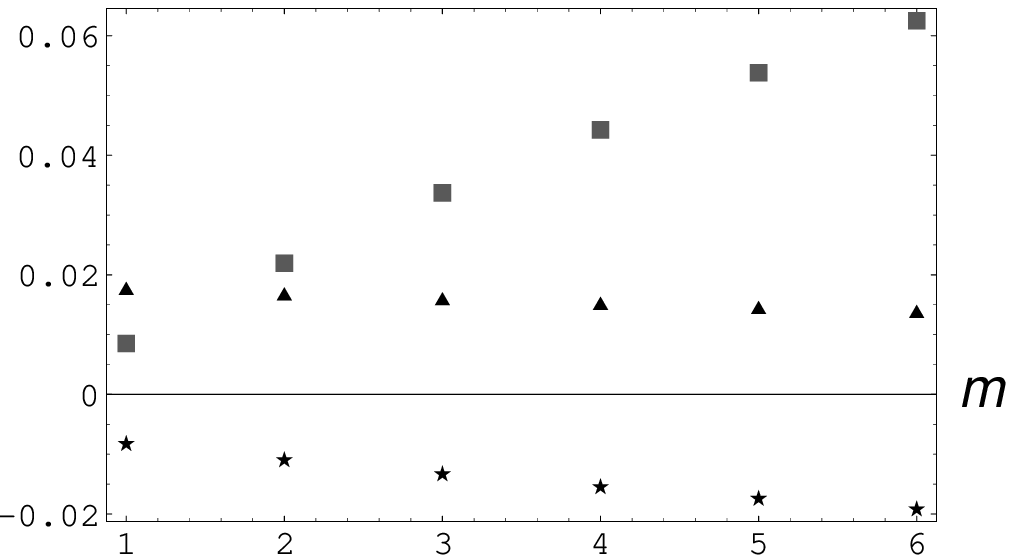, width=24em, angle=0}
\end{center}
\narrowtext
\caption{Coefficients of $\epsilon^2$ terms of the exponents $\eta_{l2}$ 
(triangles), $\eta_{l4}$ (stars), and $\beta_q$ (squares) for $n=1$.}%
\label{fig:eps2coeff}
\end{figure}

In Fig.~\ref{fig:eps2coeff} the coefficients of the $\epsilon^2$ terms
of some of these exponents are depicted for the scalar case, $n=1$.
As one sees, they have a smooth and relatively weak $m$-dependence,
especially for $\eta_{l2}$ and $\eta_{l4}$.

In the special cases $m=2$ and $m=6$, the functions $\Phi(\upsilon;m,d^*)$ and
$\Xi(\upsilon;m,d^*)$ become sufficiently simple [see (\ref{Phid*2})--(\ref{Phid*6})],
so that the required integrations can be done analytically.
This leads to 
\begin{mathletters}\label{coeffanalm2}
\begin{equation}\label{bphianalm2}
b_\phi(2)=-{1\over 54}{1\over(4\pi)^8}\;,
\end{equation}
\begin{equation}
\check{b}_\sigma(2)={1\over 162}{1\over(4\pi)^8}\;,
\end{equation}
\begin{equation}
\check{b}_\rho(2)={1\over 18}{1\over(4\pi)^8}\;,
\end{equation}
\end{mathletters}
\begin{mathletters}\label{coeffanalm6}
\begin{equation}
b_\phi(6)=-{16\over 9}\,\frac{ 1-3 \ln\!{\frac{4}{3}}}{(4\pi)^{12}}\;,
\end{equation}
\begin{equation}
\check{b}_\sigma(6)={14\over 81} {1\over (4\pi)^{12}}\;,
\end{equation}
and
\begin{equation}\label{brhoanalm6}
\check{b}_\rho(6)={8\over 9} \,\frac{ 1+6 \ln\!{\frac{4}{3}}}{(4\pi)^{12}}\;.
\end{equation}
\end{mathletters}
If these analytical expressions for the coefficients are inserted into
the expansions (\ref{etal2exp}), (\ref{etal4exp}), and (\ref{betaqexp})
of $\eta_{l2}$, $\eta_{l4}$, and $\beta_q$ with $m=2$ and $m=6$, then
Sak and Grest's\cite{SG78} results for those two values of $m$ are
recovered (which in turn agree with Mergulh{\~a}o and
Carneiro's\cite{MC99} findings for $\eta_{l2}$ and $\eta_{l4}$).

As was mentioned already in the Introduction,
these results for $m=2$ and $m=6$
{\it disagree\/} with
Mukamel's\cite{Muk77}.
More generally,  our ${\mathcal O}(\epsilon^2)$
results (\ref{nul4exp})--(\ref{betaqexp}),
for {\it all\/} values of $m=1,\ldots,6$,
turn out to be at
variance with the latter author's.
The case $m=1$ was also studied by Hornreich and Bruce\cite{HB78}, who
calculated $\eta_{l4}(m{=}1)$  and $\beta_q(m{=}1)$ to order $\epsilon^2$.
Their results agree with Mukamel's and hence diagree with our's.

Upon extrapolating the series expansions
(\ref{nul4exp})--(\ref{betaqexp}) one can obtain
exponents estimates for three-dimensional systems.
Unfortunately, there exist in the literature only very few
predictions of exponent values
produced by other means with which
we can compare our's.\cite{Sel92}
Utilizing high-temperature series techniques,
Redner and Stanley\cite{RS77} found the estimate
$\beta_q=0.5\pm 0.15$ for the case of a uniaxial $(m,d,n)=(1,3,1)$
Lifshitz point. This is in conformity with the value
$\beta_q\simeq 0.519$ one gets by setting
$\epsilon=1.5$ in the corresponding $m{=}1$ result of (\ref{betaqexp}).
A more recent high-temperature series analysis by
Mo and Ferer\cite{MF91} yielded $2\,\beta_q\simeq1$.
For the susceptibility exponent
\begin{equation}\label{gammal}
\gamma_l=\nu_{l2}(2-\eta_{l2})=\nu_{l4}(4-\eta_{l4})\:,
\end{equation}
 the correlation exponent $\nu_{l4}$, and the specific-heat exponent
\begin{equation}\label{alphal}
\alpha_l=2-m\,\nu_{l4}-(d-m)\,\nu_{l2}
\end{equation}
of the $(m,d,n){=}(1,3,1)$ Lifshitz point these authors obtained the results
$\gamma_l=1.62\pm 0.12$, $4\,\nu_{l4}=1.63\pm 0.10$,
and $\alpha_l=0.20\pm 0.15$. Utilizing these numbers to compute
$\eta_{l4}$ via the scaling law implied by
(\ref{gammal}), $\eta_{l4}=4-\gamma_l/\nu_{l4}$, yields
$\eta_{l4}\simeq 0.02\pm 0.5$. This may be be compared
with the value $\eta_{l4}\simeq -0.019$ one finds from (\ref{etal4exp})
upon setting $\epsilon=1.5$.

As a further quantity for which 
Mo and Ferer's results\cite{MF91} yield an estimate that
can be compared with our ${\mathcal O}(\epsilon^2)$ results
we consider the ratio $\beta_l/\gamma_l$. Substituting their
exponent values into $\beta_l=(2-\alpha_l-\gamma_l)/2$ yields
$\beta_l=0.09\pm0.135$ and
$\beta_l/\gamma_l=0.055^{\,-\,0.081}_{\,+\,0.094}$.
From the asymptotic form (\ref{Gsf}) of $G^{(N{=}1)}_{\mathrm ren}$
one reads off the scaling law
\begin{equation}
\beta_l=\frac{\nu_{l2}}{2}\left(d-m-2+\eta_{l2}\right)
+\frac{\nu_{l4}}{2}\,m\;,
\end{equation}
which may be combined with relation (\ref{gammal}) for $\gamma_l$ to
conclude that
\begin{equation}\label{betaovgamma}
\frac{\beta_l}{\gamma_l}={d-m-2+\eta_{l2}+m\,\frac{\nu_{l4}}{\nu_{l2}}
\over 2\,(2-\eta_{l2})}\;.
\end{equation}
We now set $m=n=1$ and $\epsilon=1.5$ in (\ref{nul4exp})
and (\ref{etal2exp}). This gives $\nu_{l4}/\nu_{l2}\simeq 0.488$ and
$\eta_{l2}\simeq 0.039$. Then we insert these numbers into
(\ref{betaovgamma}) with $d=3$, obtaining
${\beta_l}/{\gamma_l}\simeq 0.134$.

There also exist Monte Carlo estimates of exponents
for the case of a $(m,d,n){=}(1,3,1)$ Lifshitz point.\cite{Sel78,KS85}
The more recent ones, $\beta_l=0.19\pm 0.02$ and $\gamma_l=1.4\pm 0.06$,
due to Kaski and Selke\cite{KS85},
give ${\beta_l}/{\gamma_l}=0.136\pm 0.02$.
In view of the fact that
the importance of anisotropic scaling and its implications for
finite-size effects in systems exhibiting anisotropic scale
invariance\cite{BW89,Leu91} has been realized only more recently,
it is not clear to us how reliable these
Monte Carlo estimates may be expected to be.
Note, on the other hand, that the coefficients of the $\epsilon^2$ terms
of the series (\ref{nul4exp})--(\ref{betaqexp}) are all truly small.
Thus it is not unlikely that the values one gets for $d=3$
by naive evaluation of these truncated series are
fairly precise, at least for $m=1$. (The
$\epsilon^2$-corrections of these exponents grow with $m$
because of the factor $(d^*-3)^2=(1+\frac{m}{2})^2$.)

\section{Concluding remarks}\label{sec:Concl}

We have studied the critical behavior of $d$-dimensional
systems at $m$-axial Lifshitz points by means of an $\epsilon$ expansion
about the upper critical dimension $d^*=4+{m\over 2}$.
Using modern field-theory techniques, we have
been able to compute the correlation exponents $\eta_{l2}$ and $\eta_{l4}$,
the wave-vector exponent $\beta_q$, and exponents related to these via
scaling laws to order $\epsilon^2$. The resulting series expansions,
given in (\ref{nul4exp})--(\ref{betaqexp}), correct
earlier results by Mukamel\cite{Muk77} and Hornreich and
Bruce\cite{HB78}; for the special values
$m=2$ and $m=6$, we recovered Sak and Grest's\cite{SG78} findings.

To clarify this long-standing controversy, it proved useful to
work directly in position space and to compute the Laurent expansion
of the dimensionally regularized distributions associated with the Feynman
diagrams. There are two other classes
of difficult problems where this technique has demonstrated its
potential: 
field theories of polymerized (tethered) membranes\cite{DDG94,DDG97,Wie99} and critical behavior in systems with boundaries.\cite{Die86a,Die97}
In the present study an additional complication had to be mastered:
The free propagator at the Lifshitz point,
which because of anisotropic scale invariance
is  a {\it generalized homogeneous\/} function
rather than a simple power of the distance $|\boldsymbol{x}-\boldsymbol{x}'|$,
involves a complicated scaling function.
For powers and products of {\it simple\/} homogeneous functions,
a lot of mathematical knowledge on Laurent expansions is
available.\cite{GS64} Unfortunately, the amount of explicit mathematical
results on Laurent expansions of powers and products of generalized
homogeneous functions appears to be rather scarce.
Since we had no such general mathematical results at our disposal,
we had to work out the Laurent expansions of the required
distributions by our own tools.

Difficulties of the kind we were faced with in the present work
may be encountered also in studies of other types of systems with anisotropic
scale invariance. Hence the techniques utilized above
should be equally useful for field-theory analyses of such problems.
 
\acknowledgments \label{sec:Acknol}

Part of this work was done while H.~W.~D.\ was a visitor
at the physics department of Virginia Tech.\ U.
It is a pleasure to thank Beate Schmittmann, Uwe T{\"a}uber,
and Royce Zia for their warm hospitality and the pleasant atmosphere
provided during this visit.
M.\ Sh.\ would like to thank Kay Wiese
 for introducing him to the
field theory of self-avoiding membranes and
for numerous helpful discussions.
Last, but not least, we would like to express our gratitude to
Malte Henkel for comments on this work and his interest in it.

This work has been supported by the Deutsche Forschungsgemeinschaft
through the Leibniz program and Son\-der\-for\-schungs\-be\-reich 237 ``Unordnung
und grosse Fluktuationen''. 

\appendix

\section{The scaling functions $\Phi(\upsilon)$ and $\Xi(\upsilon)$}\label{App:sfPhiXi}

The scaling functions $\Phi(\upsilon)$ and
$\Xi(\upsilon)$ introduced respectively through
(\ref{Gsf})--(\ref{Phidef}) and
(\ref{Xiintro})--(\ref{Xidef})
are given
by single integrals (\ref{Phisi}) and (\ref{sfXi}) of the form
\begin{equation}\label{iv}
i(\upsilon)= \upsilon^{-\mu}\int\limits_0^\infty dq\, q^{2-\epsilon} {\mathrm J}_\mu(q\upsilon) {\mathrm K}_{\nu}(q^2)\;.
\end{equation}
This is a standard integral,\cite{PBM2}
which for arbitrary values of its parameters $\mu$ and $\nu$, can be expressed
in terms of generalized hypergeometric functions ${{_{2\!}F_{3}}}$.
For the special values $\mu={m\over 2}-1$ and
$\nu=1-{m\over 4}-{\epsilon\over 2}$ or 
$\nu=-{m\over 4}-{\epsilon\over 2}$
for which it is needed,
it simplifies, giving
\end{multicols}
\widetext\appendix\setcounter{section}{1}
\begin{equation}\label{PhiHypGeoF}
\Phi(\upsilon;m,d)={1\over 2^{2+m}\,\pi^{\frac{6 + m - 2\,\epsilon}{4}}}
\left[ {\Gamma\big(1{-}{\epsilon\over 2} \big)
\over\Gamma\big({2{+}m\over 4}\big)}
\,{{_{1\!}{\mathrm F}_{2}}}\!\left( 1{-}{\epsilon\over 2};
{1\over 2},{2{+}m\over 4};
{\upsilon^4\over 64}\right)
- {\upsilon^2\over 4}
{\Gamma\big({3\over 2}{-}{\epsilon\over 2} \big)
\over\Gamma\big(1{+}{m\over 4} \big)}
\,{{_{1\!}{\mathrm F}_{2}}}\!\left({3\over 2}{-}{\epsilon\over 2};
{3\over 2},1{+}{m\over 4};
{\upsilon^4\over 64}\right) \right]
\end{equation}
and
\begin{equation}\label{XiHypGeoF}
\Xi(\upsilon;m,d)={1\over 2^{3+m}\,
\pi^{\frac{6 + m - 2\,\epsilon}{4}}}
\left[ {\Gamma\big({1{-}\epsilon\over 2} \big)
\over\Gamma\big({m\over 4}\big)}
\,{{_{1\!}{\mathrm F}_{2}}}\!\left( {1{-}\epsilon\over 2};
{1\over 2},{m\over 4};
{\upsilon^4\over 64}\right)
- {\upsilon^2\over 4}
{\Gamma\big(1{-}{\epsilon\over 2} \big)
\over\Gamma\big({2{+}m\over 4} \big)}
\,{{_{1\!}{\mathrm F}_{2}}}\!\left(1{-}{\epsilon\over 2};
{3\over 2},{2{+}m\over 4};
{\upsilon^4\over 64}\right) \right]\,.
\end{equation}
At the upper critical dimension, i.e., for $\epsilon=0$, this
becomes
\begin{equation}\label{Phid*}
\Phi\!\Big(\upsilon;m,4{+}\frac{m}{2}\Big)=
\frac{1}{2^{5+m}\,\pi^{\frac{6{+}m}{4}}}\left[
{8\over\Gamma\big({2{+}m\over 4} \big)}
\;{{_{1\!}{\mathrm F}_{2}}}\!\left( 1;{1\over 2},{2{+}m\over 4};
{\upsilon^4\over 64}\right) -{2^{3m\over 4}\,\sqrt{\pi}}
 \,\upsilon^{2-{m\over 2}}\,
{\mathrm I}_{m\over 4}\!\Big({\upsilon^2\over 4}\Big)\right]
\end{equation}
and
\begin{equation}\label{Xid*}
\Xi\!\Big(\upsilon;m,4{+}\frac{m}{2}\Big)=
\frac{1}{2^{6+m}\,\pi^{\frac{6{+}m}{4}}}
\left[2^{3m\over 4}\,\sqrt{\pi}\,\upsilon^{2{-}{m\over 2}}\,{\mathrm I}_{{m\over 4}{-}1}
\!\Big({\upsilon^2\over 4}\Big)-
\frac{2\,\upsilon^2}{\Gamma\big({2{+}m\over 4} \big)}
\,{{_{1\!}{\mathrm F}_{2}}}\!\left(1;
{3\over 2},{2{+}m\over 4};
{\upsilon^4\over 64}\right)
\right]
\;,
\end{equation}
respectively,
where the ${\mathrm I}_{\nu}(.)$ are modified Bessel functions of the first kind.

\begin{multicols}{2}
In the special cases $m=2$ and $m=6$, these expressions
reduce to simple elementary functions: One has
\begin{equation}\label{Phid*2}
\Phi(\upsilon;2,5)={1\over (4 \pi)^2}\,{\mathrm{e}}^{-{\upsilon^2\over 4}}\;,
\end{equation}
\begin{equation}
\Xi(\upsilon;2,5)={1\over 2}\,\Phi(\upsilon;2,5)\,,
\end{equation}\label{Xid*2}
\begin{equation}\label{Phid*6}
\Phi(\upsilon;6,7)={1-(1+{\upsilon^2\over 4})\,
{\mathrm{e}}^{-{\upsilon^2\over 4}}\over (2 \pi)^3\,\upsilon^4}\;,
\end{equation}
and
\begin{equation}\label{Xid*6}
\Xi(\upsilon;6,7)={1\over (4 \pi)^3}\,{1\over\upsilon^2}\left ( 1-{\mathrm{e}}^{-{\upsilon^2\over 4}}\right )\;.
\end{equation}

The reason for the latter simplifications is the following.
If $m=2$ or $m=6$ and  $d=d^*=4+m/2$ (upper critical dimension),
then Bessel functions ${\mathrm K}_\nu$ with $\nu=\pm\frac{1}{2}$
are encountered in the integral  (\ref{iv}), which are simple
exponentials.\cite{SFSimpl} This entails that
the required single integrations
can be done analytically to obtain the results
(\ref{bphianalm2})--(\ref{brhoanalm6}) for the ${\mathcal O}(\epsilon^2)$
coefficients.

For the remaining values of $m$, i.e., for $m=1,3,4,5$,
the required integrals did not simplify to a  degree that we were able to compute them
analytically.  However, proceeding as explained in Appendix \ref{App:AsBeh},
they can be computed numerically.
In the special cases $m=2$ and $m=6$, the results of our numerical integrations
are in complete conformity with the analytical ones.

\section{Asymptotic behavior of the scaling functions
$\Phi(\upsilon)$ and $\Xi(\upsilon)$}\label{App:AsBeh} 

According to (\ref{Phid*}), the scaling function $\Phi(\upsilon;m,d^*)$
is a difference of a hypergeometric function ${{_{1\!}{\mathrm F}_{2}}}$
and a product of a Bessel function ${\mathrm I}_{m/4}$ times a power.
If one asks Mathematica\cite{MAT} to numerically evaluate
expression (\ref{Phid*}) for $\Phi(\upsilon)$ without taking
any precautionary measures, the result becomes
inaccurate whenever $\upsilon$ becomes sufficiently large.
We found that such a direct, naive numerical evaluation
fails for values of $\upsilon$ exceeding $\upsilon_0\simeq 9.5$.
This is because both functions of this difference increase exponentially
as $\upsilon\to\infty$.

To cope with this problem, we determined the asymptotic
behavior of the scaling functions
$\Phi(\upsilon;m,d^*)$ and $\Xi(\upsilon;m,d^*)$ for $\upsilon\to\infty$.
From the integral representations (\ref{Phidef}) and
(\ref{Xidef}) of these functions one easily derives the limiting
forms
\begin{equation}\label{Phias}
\Phi(\upsilon;m,d)
\mathop{\approx}\limits_{\upsilon\to \infty}
\Phi^{({\mathrm as})}(\upsilon;m,d)\equiv
 \upsilon^{-4+2 \epsilon}\,\Phi_\infty(m,d)
\end{equation}
and
\begin{equation}\label{Xias}
\Xi(\upsilon;m,d)
\mathop{\approx}\limits_{\upsilon\to \infty}
\Xi^{({\mathrm as})}(\upsilon;m,d)\equiv
 \upsilon^{-2+2 \epsilon}\,{\Phi_\infty(m,d)\over 8\,(1-\epsilon)}\;,
\end{equation}
with
\begin{eqnarray}\label{Phiinf}
\Phi_\infty(m,d)&=&\int\limits_{\boldsymbol{q}_\|}
\int\limits_{\boldsymbol{q}_\perp}
\frac{{\mathrm{e}}^{i(\boldsymbol{q}_\parallel\cdot\boldsymbol{e}_\parallel)}}
{q_\parallel^4+q_\perp^2}\nonumber\\
&=&\frac{2^{ 2(d - m)-5}\,\pi^{\frac{1 - d}{2}}\,
\Gamma\big( d{-}2 {-} \frac{m}{2}\big)}{
\Gamma \big(\frac{3 {-} d {+} m}{2}\big)}\;.
\end{eqnarray}
At the upper critical dimension, the latter coefficient
becomes
\begin{equation}\label{Phiinfd*}
\Phi_{\infty}(m,d^*)={2^{3-m}\,\pi^{-{6+m\over 4}}\,
\over{
\Gamma\big({m-2\over 4}\big)}}\,.
\end{equation}
Note that for $m=2$
the asymptotic form (\ref{Phias}) is consistent with
the simple exponential form (\ref{Phid*2})
since $\Phi_{\infty}(2,5)=0$. However, for other values
of $m$, the coefficient (\ref{Phiinfd*}) does {\it not\/} vanish.
For example, $\Phi_{\infty}(6,7)=1/(8\,\pi^3)$,
in conformity with expression (\ref{Phid*6}) for the scaling function $\Phi(\upsilon;6,7)$.

In order to obtain precise results for the integrals
$J_{0,3}(m,d^*)$, $J_{4,3}(m,d^*)$, and $I_1(m,d^*)$,
on which the coefficients $b_\phi(m)$, $\check{b}_\sigma(m)$, and
$\check{b}_\rho(m)$ depend, we proceeded as follows.
We split the required integrals as
$\int_0^\infty \ldots d\upsilon=\int_0^{\upsilon_0}\ldots d\upsilon
+\int_{\upsilon_0}^\infty \ldots d\upsilon$, choosing $\upsilon \simeq 9.3$.
In the  integrals $\int_{\upsilon_0}^\infty \ldots d\upsilon$,
we replaced the integrands
by their asymptotic forms obtained upon substitution of $\Phi$ and/or
$\Xi$ by their large-$\upsilon$ approximations
$\Phi^{({\mathrm as})}$ and $\Xi^{({\mathrm as})}$ given in (\ref{Phias})
and (\ref{Xias}), respectively, and then computed these integrals
analytically. The integrals $\int_0^{\upsilon_0}\ldots d\upsilon$
were computed numerically, using Mathematica.\cite{MAT}
We checked that reasonable changes of $\upsilon_0$
have negligible effects on the results.
The procedure yields very accurate numerical values of the
requested integrals. 
The reader may convince himself of the precision by comparing the
so-determined numerical values of the integrals
for $m=2$ and $m=6$ with the
analytically known exact values.

\end{multicols}

\begin{thebibliography}{10}
\item[$^\ddag$]Permanent address.
\item[$^\dag$] Permanent address:
Institute for
Condensed Matter Physics, 1 Svientsitskii str, 79011 Lviv, Ukraine

\bibitem{HLS75b}
R.~M. Hornreich, M. Luban, and S. Shtrikman, Phys. Rev. Lett. {\bf 35},  1678
  (1975).

\bibitem{HLS75}
R.~M. Hornreich, M. Luban, and S. Shtrikman, Phys. Lett. {\bf 55A},  269
  (1975).

\bibitem{Hor80}
R.~M. Hornreich, J. Magn. Magn. Mat. {\bf 15--18},  387  (1980).

\bibitem{Sel92}
W. Selke,  in {\em Phase Transitions and Critical Phenomena}, edited by C. Domb
  and J.~L. Lebowitz (Academic Press, London, 1992), Vol.~15, pp.\ 1--72.

\bibitem{SBOC81}
Y. Shapira, C. Becerra, N.~O. Jr., and T. Chang, Phys. Rev. B {\bf 24},  2780
  (1981).

\bibitem{FurRef}
For a more complete set of references, see Refs.\ \onlinecite{Hor80} and
  \onlinecite{Sel92}.

\bibitem{VS92}
Y.~M. Vysochanski\u{\i} and V.~Y. Slivka, Usp. Fiz. Nauk {\bf 162},  139
  (1992), [{S}ov. Phys. Usp. {\bf 35}, 123--134].

\bibitem{AD77}
E. Abraham and I.~E. Dzyaloshinskii, Solid State Commun. {\bf 23},  883
  (1977).

\bibitem{HZW80}
C. Hartzstein, V. Zevin, and M. Weger, J. Phys. (France) I {\bf 41},  677
  (1980).

\bibitem{CL76}
J.~H. Chen and T.~C. Lubensky, Phys. Rev. A {\bf 14},  1202  (1976).

\bibitem{AM80}
A. Aharony and D. Mukamel, J. Phys. C {\bf 13},  L255  (1980).

\bibitem{LN79}
J. Lajzerowicz and J.~J. Niez, J. Phys. (Paris) Lett. {\bf 40},  L165  (1979).

\bibitem{FS80}
M.~E. Fisher and W. Selke, Phys. Rev. Lett. {\bf 44},  1502  (1980).

\bibitem{FS81}
M.~E. Fisher and W. Selke, Phil. Trans. R. Soc. Lond. {\bf 302},  1  (1981).

\bibitem{Gum86}
G. Gumbs, Phys. Rev. B {\bf 33},  6500  (1986).

\bibitem{BF99}
K. Binder and H.~L. Frisch, Eur. Phys. J. {\bf 10},  71  (1999).

\bibitem{FKB00}
H.~L. Frisch, J.~C. Kimball, and K. Binder, J. Phys. C {\bf 12},  29  (2000).

\bibitem{HH77}
B.~I. Halperin and P.~C. Hohenberg, Rev. Mod. Phys. {\bf 49},  435  (1977).

\bibitem{SZ95}
B. Schmittmann and R.~K.~P. Zia, {\em Statistical Mechanics of Driven Diffusive
  Systems}, Vol.~17 of {\em Phase Transitions and Critical Phenomena} (Academic
  Press, London, 1995), pp.\ 1--220.

\bibitem{Kru97}
J. Krug, Adv. Phys. {\bf 46},  139  (1997).

\bibitem{ASIERW}
For example, in a recent investigation\cite{Hen97} of the question of whether
  anisotropic scale invariance may lead to an even larger invariance group
  (reminiscent of conformal or Schr{\"o}dinger invariance) such systems were
  explicitly considered.

\bibitem{Hen97}
M. Henkel, Phys. Rev. Lett. {\bf 78},  1940  (1997).

\bibitem{Muk77}
D. Mukamel, J. Phys. A {\bf 10},  L249  (1977).

\bibitem{HB78}
R.~M. Hornreich and A.~D. Bruce, J. Phys. A {\bf 11},  595  (1978).

\bibitem{SG78}
J. Sak and G.~S. Grest, Phys. Rev. B {\bf 17},  3602  (1978).

\bibitem{GS78}
G.~S. Grest and J. Sak, Phys. Rev. B {\bf 17},  3607  (1978).

\bibitem{NTCS76}
J.~F. Nicoll, G.~F. Tuthill, T.~S. Chang, and H.~E. Stanley, Phys. Lett. A {\bf
  58},  1  (1976).

\bibitem{IHB90}
A.~A. Inayat-Hussain and M.~J. Buckingham, Phys. Rev. A {\bf 41},  5394
  (1990).

\bibitem{MC98}
C. Mergulh{\~a}o, Jr. and C.~E.~I. Carneiro, Phys. Rev. B {\bf 58},  6047
  (1998).

\bibitem{MC99}
C. Mergulh{\~a}o, Jr. and C.~E.~I. Carneiro, Phys. Rev. B {\bf 59},  13954
  (1999).

\bibitem{GS64}
I.~M. Gel'fand and G.~E. Shilov,  in {\em Generalized Functions} (Academic
  Press, New York and London, 1964), Vol.~1, Chap.~3.9 and 4.6.

\bibitem{Die86a}
H.~W. Diehl,  in {\em Phase Transitions and Critical Phenomena}, edited by C.
  Domb and J.~L. Lebowitz (Academic Press, London, 1986), Vol.~10, pp.\
  75--267.

\bibitem{J02calc}
To see this within the present calculational scheme, note that
  $J_{0,2}(m,d)/S_m$ is nothing but the Fourier transform of $\Phi^2(\upsilon)$
  at zero momentum $\boldsymbol{q}_\|$. Hence it is equal to the integral
  $\int_{\boldsymbol{q}_\|}$ over the square of the Fourier transform of
  $\Phi(\upsilon)$, which can be read off from the Fourier-integral
  representation (\ref{PhiF}) of $\Phi(\upsilon)$. Upon performing the required
  angular integrations, one is left with a standard integral, which yields
  (\ref{J02}).

\bibitem{RS77}
S. Redner and H.~E. Stanley, Phys. Rev. B {\bf 16},  4901  (1977).

\bibitem{MF91}
Z. Mo and M. Ferer, Phys. Rev. B {\bf 43},  10890  (1991).

\bibitem{Sel78}
W. Selke, Z. Phys. {\bf 29},  133  (1978).

\bibitem{KS85}
K. Kaski and W. Selke, Phys. Rev. B {\bf 31},  3128  (1985).

\bibitem{BW89}
K. Binder and J.-S. Wang, J. Stat. Phys. {\bf 55},  87  (1989).

\bibitem{Leu91}
K.-t. Leung, Phys. Rev. Lett. {\bf 66},  453  (1991).

\bibitem{DDG94}
F. David, B. Duplantier, and E. Guitter, Phys. Rev. Lett {\bf 72},  311
  (1994).

\bibitem{DDG97}
F. David, B. Duplantier, and E. Guitter, preprint cond-mat/9702136
  (unpublished).

\bibitem{Wie99}
K. Wiese, Habilitationsschrift, U. Essen, 1999;
 in {\em Phase Transitions and Critical Phenomena}, edited by C.
  Domb and J.~L. Lebowitz (Academic Press, London, 2000), Vol.~19.

\bibitem{Die97}
H.~W. Diehl, Int.\ J.\ Mod.\ Phys.\ B {\bf 11},  3503  (1997), preprint
  cond-mat/9610143.

\bibitem{PBM2}
A. Prudnikov, Y.~A. Brychkov, and O.~I. Marichev, {\em Integrals and Series}
  (Gordon and Breach, New York, 1986), Vol.~2, eq. 2.16.22.6.

\bibitem{SFSimpl}
Both Sak and Grest \cite{SG78} as well as Mergulh{\~a}o and Carneiro
  \cite{MC99} explicitly employed the simple form (\ref{Phid*2}) of the scaling
  function in their computations. However, the latter authors \cite{MC99} did
  not take advantage of working with the simple function (\ref{Phid*6}) in the
  case $m=6$, performing complicated computations in the momentum
  representation instead. Sak and Grest\cite{SG78}, on the other hand, did not
  present any details of their calculation for the case $m=6$.

\bibitem{MAT}
Mathematica, version 3.0, a product of Wolfram Research.

\end{thebibliography}
\end{document}